\documentclass[12pt]{article}
\pdfoutput=1
\usepackage{amssymb,amsmath}
\usepackage{graphicx}
\usepackage{color}
\usepackage[colorlinks=true
,urlcolor=blue
,citecolor=blue
,linkcolor=blue
,pagecolor=blue
,linktocpage=true
,pdfproducer=medialab
]{hyperref}
\usepackage[a4paper]{geometry}

\usepackage{cite}
\usepackage{placeins}
\usepackage{slashed}
\usepackage{mathtools}
\usepackage[FIGTOPCAP]{subfigure}

\makeatletter \renewcommand{\@dotsep}{10000} \makeatother

\mathchardef\mhyphen="2D

\newcommand{\beq}{\begin{equation}}
\newcommand{\eeq}{\end{equation}}
\newcommand{\bea}{\begin{eqnarray}}
\newcommand{\eea}{\end{eqnarray}}

\newcommand{\mgut}{M_{\rm GUT}}

\usepackage[nodisplayskipstretch]{setspace}

\begin{document}

\begin{titlepage}
\pagestyle{empty}

\vspace*{0.2in}
\begin{center}
{\Large \bf  Stop Search in SUSY SO(10) GUTs with Nonuniversal Gaugino Masses}\\
\vspace{1cm}
{\bf  Zafer Alt\i n$^a$\footnote{E-mail: 501407009@ogr.uludag.edu.tr},
Zerrin K\i rca$^a$\footnote{E-mail: zkirca@uludag.edu.tr},
Tu\~{g}\c{c}e Tan\i mak$^a$\footnote{E-mail: 501807001@ogr.uludag.edu.tr}
 and
Cem Salih $\ddot{\rm U}$n$^a$\footnote{E-mail: cemsalihun@uludag.edu.tr}}
\vspace{0.5cm}

{\it $^a$Department of Physics, Bursa Uluda\~{g} University, TR16059 Bursa, Turkey} \\

\end{center}

\vspace{0.5cm}
\begin{abstract}
\noindent
We have discussed the stop mass and possible signal processes within a class of SUSY GUTs with non-universal gaugino masses. This class of models predicts the stop mass in a wide range  from about 400 GeV to 8 TeV, and the DM constraints bound it as $m_{\tilde{t}_{1}}\gtrsim 500$ GeV.  Being the lightest supersymmetric particle the neutralino always takes part in possible signal processes, and its mass is realized as heavy as about 2.3 TeV in the fundamental parameter space. Similarly, the lightest chargino mass can be realized beyond 3 TeV, while the DM constraints bound its mass at about 2.7 TeV from above. We find that the stop mass below about 1.2 TeV is excluded by the analyses over the $\tilde{t}_{1}\rightarrow t\tilde{\chi}_{1}^{0}$ decay mode performed under the current experimental setups. This mode can help probe the stop mass up to about 6 TeV in future collider experiments. Similar analyses yield that the stop mass will be able to be probed to about 4.8 TeV and 5 TeV, if the $\tilde{t}_{1}\rightarrow b W^{\pm}\tilde{\chi}_{1}^{0}$ or $\tilde{t}_{1}\rightarrow b q\bar{q}^{\prime}\tilde{\chi}_{1}^{0}$ decay modes are allowed. We show that the former decay mode is not available for this class of SUSY GUTs in the current experiments, while it will be able to be tested in future.

\end{abstract}

\end{titlepage}

\setcounter{footnote}{0}

\section{Introduction}
\label{ch:introduction}

The Standard Model (SM) of the elementary particles is one of the most successful theory in physics, and its glory was embraced especially after the Higgs boson discovery by the ATLAS \cite{Aad:2012tfa} and CMS \cite{Chatrchyan:2013lba} experiments. Despite its success and good agreement between its predictions and the experimental results, the SM can only be an effective theory due to its problematic structure for the Higgs boson. The most significant problem arises in stabilizing the Higgs boson mass against the quadratic divergent radiative contributions, which is very well-known as the gauge hierarchy problem \cite{Gildener:1976ai}. Besides, the SM with a 125 GeV Higgs boson loses the absolute stability of the Higgs potential \cite{hinstability} at some high scale, which may arise another reason for the need of models beyond the SM. As one of the forefront candidate models, Supersymmetry (SUSY) can resolve the gauge hierarchy problem by extending the SM with superpartners such that the divergent contributions to the Higgs boson mass from the SM particles are canceled by those from their superpartners. Besides, imposing the conservation of $R-$parity requires the lightest supersymmetric particle (LSP) to be stable, and neutral weakly interacting SUSY particles can provide pleasant dark matter (DM) candidates. In addition, the minimal supersymmetric extension of the SM (MSSM) unifies the three SM gauge couplings, even though it preserves the SM gauge symmetry. Together with stabilizing the Higgs boson mass at all the energy scales, the gauge coupling unification motivates the supersymmetric grand unified theories (SUSY GUTs), and one can explore low scale implications of GUTs by linking the high scale origin to the low energy observables through the renormalization group equations (RGEs), which has allowed to explore the implications of GUTs based on SU(5)\cite{Chattopadhyay:2001mj} or SO(10) \cite{big-422,bigger-422}. 

Even though the experimental observations and constraints from the Higgs boson searches (see, for instance, \cite{ATLAS:2019aqa,ATLAS:2018uoi,Khachatryan:2015cwa,Sirunyan:2017dgc,Sirunyan:2018kst,Aad:2019zwb,Sirunyan:2018wim,Ko:2016asv,Khachatryan:2015kon}) point a need for new physics, and the observations can be accommodated in SUSY GUTs \cite{Abdallah:2018kix}, absence of a direct signal in the experiments brings a strong impact in searches for the new physics. As is a hadron collider, the LHC results are quite effective, especially on new colored particles such as gluino and stop in the SUSY models. The current LHC results have excluded the gluino lighter than about 2.4 TeV, while it reduces to about 2.2 TeV as the LSP neutralino mass decreases \cite{Aaboud:2017vwy}. Moreover, the exclusion reduces as $m_{\tilde{g}}\gtrsim 800$ GeV, if the gluino happens to be next to LSP (NLSP). A recent study has shown that these bounds can also be employed in the low scale mass spectra of SUSY GUTs \cite{Altin:2019veq}.

Contributing through RGEs heavy gluino mass scales exclude the stop solutions lighter than about 400 GeV in the MSSM framework \cite{Altin:2019veq}. A further exclusion on the stop mass can be obtained by performing collider analyses over the possible decay modes of the stop. The most stringent exclusion on the stop mass arises, if the stop decays into a top quark and a LSP neutralino. Results from the ATLAS and CMS experiments \cite{Aaboud:2017vwy} exclude the solutions with $m_{\tilde{t}_{1}} \lesssim 1200$ GeV. Formation of the LSP also takes part in analyses such that if the Higgsinos take part in the formation of the LSP, then the exclusion happens as $m_{\tilde{t}_{1}}\gtrsim 900$ GeV. \cite{Mitrevski:2018iyy}. Besides, the number of leptons in the final states can lower the exclusion. While the expected exclusion arises up to about 1200 GeV for the stop mass, the analyses of the decay processes with final states of two leptons yield a lower exclusion as $m_{\tilde{t}_{1}}\gtrsim 800$ GeV \cite{Collard:2017yec}. Similar results can be obtained if the stop is allowed to decay into a chargino along with a bottom quark, which resumes on the chargino decay into a LSP neutralino and a $W-$boson. The analyses over such events exclude the solutions with $m_{\tilde{t}_{1}} \lesssim 1100$ GeV \cite{Aaboud:2017vwy,ATLAS:2016jaa}. The lowest bound on the stop mass is obtained when the stop can decay only into a LSP neutralino along with a charm quark. This decay mode is not very exclusive due to the soft charm jets and low missing energy \cite{Ajaib:2011hs}. Such events exclude the stop mass as $m_{\tilde{t}_{1}} \lesssim 550$ GeV \cite{Sirunyan:2017wif}.

Most of these analyses, on the other hand, are mostly performed in the low scale SUSY, and they assume specific configurations based on the mass differences of the relevant particles, 100\% branching ratios for the decay modes under concern, largest production cross-sections etc. In addition, the rest of the SUSY mass spectrum is assumed to be heavy enough not to interfere in the events. Such configurations are quite common and they can be easily adjusted, since the fundamental parameter space of the low scale SUSY is spanned by more than one hundred free parameters including masses, mixings, soft supersymmetry breaking (SSB) trilinear couplings and so on. On the other hand, if one considers the low scale implications of a class of SUSY GUTs, then some of the configurations may not be possible, since a quite large number of observables are calculated in terms of a few free parameters. For instance, when the stop is allowed to decay into a top quark and a LSP neutralino, a class of SUSY GUTs with non-universal gaugino masses (NUGM) at the grand unification scale ($\mgut$) can yield an exclusion at about 50\% Confidence Level (CL), at most, in the region with $m_{\tilde{t}_{1}}\lesssim 500$ GeV (see, for instance, Ref.\cite{Demir:2014jqa}), even though the low scale analyses exclude the same region at 95\% CL \cite{Mitrevski:2018iyy, Collard:2017yec}.

In this work, we simulate the similar analyses of signal-background comparisons in the framework of SUSY GUTs with non-universal gaugino masses at the GUT scale. The rest of the paper is organized as follows: Section \ref{sec:scan} describes the scanning procedure and the experimental constraints employed in our analyses. Also we define the signal strength and confidence level (CL) in this section. We discuss the mass spectrum and the availability of the signal processes in Section \ref{sec:mass}, and we quantitatively present our results for the exclusion curves for the stop mass scales in the current experiments in Section \ref{sec:stopprobe}. We also show our results for probing the stop mass in future experiments through the possible signal processes in this section. Finally, we summarize and conclude our findings in Section \ref{sec:conc}.

\section{Scanning Procedure and Experimental Constraints}
\label{sec:scan}

We have employed SPheno 4.0.3 package \cite{Porod:2003um, Porod2} {generated} with SARAH 4.13.0 \cite{Staub:2008uz,Staub2}. In this package, the weak scale values of the gauge and Yukawa couplings in MSSM are evolved to the unification scale $M_{{\rm GUT}}$ via the renormalization group equations (RGEs). $M_{{\rm GUT}}$ is determined by the requirement {of unification of the gauge couplings} through their RGE evolutions. Note that we do not strictly enforce the unification condition $g_1 = g_2 = g_3$ at $M_{{\rm GUT}}$ since a few percent deviation from the unification can be assigned to unknown GUT-scale threshold corrections \cite{Hisano:1992jj,GUTth}. Afterwards, the boundary conditions are implemented at $M_{{\rm GUT}}$ and all the soft supersymmetry breaking (SSB) parameters along with the gauge and Yukawa couplings are evolved back to the weak scale. 

We performed random scans in the parameters space of NUGM, which is spanned by the following parameters;

\begin{equation}
\begin{array}{ccc}
 0  \leq & m_{0} & \leq 5~{\rm TeV}~, \\
 0 \leq & M_{1},M_{2},M_{3} & \leq 5~{\rm TeV}~, \\
-3 \leq & A_{0}/m_{0} & \leq 3~, \\
1.2 \leq & \tan\beta & \leq 60~.
\end{array}
\label{paramSP}
\end{equation}
where $m_0$ is the universal SSB mass term for the matter scalars and Higgs fields. $M_{1}$, $M_{2}$ and $M_{3}$ are the SSB mass terms for the gauginos associated with the  $U(1)_{Y}$, $SU(2)_{L}$ and $SU(3)_{C}$ symmetry groups respectively. $A_0$ is {the} SSB trilinear coupling, and $\tan\beta$ is ratio of vacuum expectation values (VEVs) of the MSSM Higgs doublets. Finally, we have used {the} central value of top quark mass, $m_t=173.3$ GeV \cite{Group:2009ad}. Note that the sparticle spectrum is not too sensitive {for} one or two sigma variation in the top quark mass \cite{Gogoladze:2011db}, but it can shift the Higgs boson mass by  1-2 GeV \cite{Gogoladze:2011aa,Ajaib:2013zha}.

The REWSB condition provides a strict theoretical constrain \cite{Ibanez:Ross,REWSB2,REWSB3,REWSB4,REWSB5} over the fundamental parameter space given in Eq.(\ref{paramSP}). The $\mu-$term is determined by the radiative electroweak symmetry breaking (REWSB) condition through its square, but its sign remains undetermined. We assume it to be positive, and accept only solutions which are compatible with the REWSB condition in our scans. Another strong  constraint comes from the relic abundance of charged supersymmetric particles \cite{Nakamura:2010zzi}. This constraint excludes regions which yield charged particles such as stop and stau to be the LSP. In this context, we require the solutions to yield one of the neutralinos to be the LSP. {In this case}, it is also {appropriate} that the LSP {becomes a suitable dark matter candidate}. {The thermal} relic abundance of LSP should, {of course}, be consistent with the current results from the WMAP \cite{Hinshaw:2012aka} and Planck \cite{Akrami:2018vks} satellites. However, even if a solution does not satisfy the dark matter observations, it can still survive in conjunction with other form(s) of the  dark matter \cite{Baer:2012by}. {We} mostly focus on the LHC allowed solutions, {but} we also discuss the DM implications in our results.

In addition to these requirements, we also successively apply the mass bounds on the supersymmetric mass spectrum  \cite{Agashe:2014kda} and the constraints from the rare B-decays ($B_s \rightarrow \mu^+ \mu^-$ \cite{Aaij:2012nna}, $B_s \rightarrow X_s \gamma$ \cite{Amhis:2012bh} and $B_u \rightarrow \tau \nu_\tau $ \cite{Asner:2010qj}). In applying the mass bounds, we listed the Higgs boson \cite{Aad:2012tfa,Chatrchyan:2013lba} and gluino \cite{Aad:2019ftg} masses separately, since they have received further updates by the LHC experiments since LEP II. These constraints can be listed as follows:

\begin{equation}
\setstretch{2.0}
\begin{array}{l}
123 \leq m_{h} \leq 127~{\rm GeV}\\
2100~{\rm GeV} \leq m_{\tilde{g}}  \\
 0.8\times 10^{-9} \leq BR (B_s \rightarrow {\mu}^{+} {\mu}^{-}) \leq 6.2 \times 10^{-9}~ (2\sigma) \\
2.9\times 10^{-4} \leq BR (b \rightarrow s {\gamma})\leq 3.87\times 10^{-4}~ (2\sigma)\\
0.15 \leq \dfrac{BR (B_u \rightarrow {\nu}_{\tau} {\tau})_{MSSM}}{BR (B_u \rightarrow {\nu}_{\tau} {\tau})_{SM}}\leq 2.41 ~ (3\sigma) \\
0.0913 \leq \Omega h^{2}({\rm WMAP}) \leq 0.1363 ~(5\sigma) \\
0.114 \leq \Omega h^{2}({\rm Planck}) \leq 0.126 ~(5\sigma)~.
\end{array}
\label{constraints}
\end{equation}
Note that we have employed the mass bound on the gluino as $m_{\tilde{g}}\geq 2.1$ TeV to take the uncertainty of about 10\% in calculation of the mass spectrum into account. 

In scanning the parameter space we use {an} interface, which employs {the} Metropolis-Hasting algorithm described in  \cite{Belanger:2009ti,SekmenMH}. After generating the low scale data with SPheno, all outputs are transferred to {MicrOmegas} \cite{Belanger:2006is} for calculations of the relic abundance of the LSP neutralino as a candidate for DM.  The solutions satisfying all the constraints mentioned above (except those from the DM observations) are so-called LHC allowed solutions. Once they are obtained, we transferred their output files from SPheno to MadGraph \cite{Alwall:2011uj} for calculation of the cross-sections for the possible signal processes and relevant SM backgrounds. The possible signal processes to detect the stop can be summarized as follows \cite{Mitrevski:2018iyy}:

\begin{equation}
\setstretch{2.5}
\begin{array}{cl}
{\rm Signal~1}: & p p \rightarrow \tilde{t}_{1}\tilde{t}_{1} \xrightarrow{\text{\large $\tilde{t}_{1}\rightarrow t\tilde{\chi}_{1}^{0}$}}t\bar{t}\tilde{\chi}_{1}^{0}\tilde{\chi}_{1}^{0} \\
{\rm Signal~2}: & p p \rightarrow \tilde{t}_{1}\tilde{t}_{1} \xrightarrow{\text{\large $\tilde{t}_{1}\rightarrow b\tilde{\chi}_{1}^{\pm}$}}b\bar{b}\tilde{\chi}_{1}^{\pm}\tilde{\chi}_{1}^{\mp} \xrightarrow{\text{\large $\tilde{\chi}_{1}^{\pm}\rightarrow W^{\pm}\tilde{\chi}_{1}^{0}$}}\rightarrow b\bar{b}W^{\pm}W^{\mp} \tilde{\chi}_{1}^{0}\tilde{\chi}_{1}^{0} \\
{\rm Signal~3}: & p p \rightarrow \tilde{t}_{1}\tilde{t}_{1} \xrightarrow{\text{\large $\tilde{t}_{1}\rightarrow b\tilde{\chi}_{1}^{\pm}$}}b\bar{b}\tilde{\chi}_{1}^{\pm}\tilde{\chi}_{1}^{\mp} \xrightarrow{\text{\large $\tilde{\chi}_{1}^{\pm}\rightarrow q\bar{q}^{\prime}\tilde{\chi}_{1}^{0}$}}\rightarrow b\bar{b}(q\bar{q}^{\prime})(q\bar{q}^{\prime}) \tilde{\chi}_{1}^{0}
\end{array}
\label{eq:sigs}
\end{equation}

The signal processes require the mass difference between the lightest stop and the LSP neutralino to be greater than the mass of a top quark (for Signal 1) or a $W-$boson approximately (for Signal 2). If the stop is not realized to be heavy enough, then the stop may involve in the processes in which it decays into a charm quark along with a LSP neutralino. However, the charm quark signal is overwhelmed by the process $\tilde{t}_{1}\rightarrow f\bar{f}^{\prime}b\tilde{\chi}_{1}^{0}$ \cite{Muhlleitner:2011ww}, and such a signal cannot provide a clear detection for the stop, and it can probe the stop mass up to about 230 GeV \cite{TheATLAScollaboration:2013aia}. Even though a similar discussion can be followed for Signal 2, the mass difference between the stop and LSP neutralino is not the only factor, but the formations of the stop and the chargino are also important. In the case of Wino-like chargino, the lightest stop should be formed mostly by the left-handed stop, since the chirality of $SU(2)_{L}$ forbids the Wino to interact with the right-handed particles. On the other hand, if the Higgsino forms the lightest chargino, then the decay mode is open for both left- and right-handed stops. Moreover, the signal is expected to be stronger when Higgsino takes part, since the interaction is proportional to the top-quark Yukawa coupling and trilinear scalar interaction term $A_{t}$. However, a Higgsino-like chargino leads to another suppression from the $\tilde{\chi}_{1}^{\pm}\rightarrow W^{\pm}\tilde{\chi}_{1}^{0}$. The chargino-neutralino-W vertex in MSSM can be written as

\begin{equation}
\setstretch{2.0}
\begin{array}{rl}
\Gamma_{\tilde{\chi}_{1}^{\pm}-\tilde{\chi}_{1}^{0}-W_{\mu}}= & -\dfrac{i}{2}g_{2}\left( 2U^{*}_{11}N_{12}+\sqrt{2}U^{*}_{12}N_{13}  \right)\left(\gamma_{\mu}\dfrac{1-\gamma_{5}}{2}   \right) \\ 
& + \dfrac{i}{2}g_{2}\left( \sqrt{2}N^{*}_{14}V_{12}-2N^{*}_{12}V_{11} \right)\left(\gamma_{\mu}\dfrac{1+\gamma_{5}}{2}   \right)
\end{array}
\end{equation}
where $N_{ij}$ is the matrix coding the mixing of bino, wino and Higgsinos in formation of neutralino mass eigenstates, while $U_{ij}$ and $V_{ij}$ diagonalize the chargino mass matrix. If the higgsino dominates in the lightest chargino formation ($U_{12} (V_{12})\gg U_{11} (V_{11})$), a strong signal also requires $N_{13}$ and $N_{14}$ to dominate in the LSP neutralino formation. In this case, the Higgsinos form both the LSP neutralino and lightest chargino, and so both are of mass about $\mu$, which kinematically forbids $\tilde{\chi}_{1}^{\pm}\rightarrow W^{\pm}\tilde{\chi}_{1}^{0}$. Another possibility is that the LSP neutralino can be formed mostly by bino, while the lightest chargino remains higgsino-like. In this case $N_{13}$ and $N_{14}$ will be small, and the signal might not still yield any visible effect. In sum, Signal 2 is available if the lightest stop mass eigenstate is formed mostly by the left-handed stop. However, a similar signal can be realized when the right-handed stop takes part in the processes. In this case, the chargino decay into two quarks along with a LSP neutralino can be considerable, which is shown in the decay cascades of Signal 3.

All the possible signal processes start with the production of a pair of stop quarks; which occurs through $q\bar{q}$ and $gg$ interactions. $q\bar{q}$ interactions can contribute when the stop is light. In both cases, the stop pair-production is mediated by the gluon; thus, the cross-section can be calculated by considering the virtual and soft gluon processes, as well as hard gluon processes. While all these processes can contribute either positively or negatively, the processes involving $q\bar{q}$ interactions and/or hard gluon contributions become negligible at the heavy mass scales of the stop. When the stop weighs more than about 1 TeV, the gluon interactions with virtual and soft gluon mediators are expected to be the main channel in the stop pair-production (for more details, see \cite{Beenakker:1997ut}).

Since the stop pair-production is the trigger in all the signal processes, the pair production of the top quarks forms the most dominant background processes. Nevertheless, its final state is quite similar to the signal processes, and the cuts suppressing the background lead to a significant decrease in the signal cross-section \cite{Demir:2014jqa}. Before concluding, we should note that the following approximation has been used in calculation of the cross-sections:

\begin{equation}
\setstretch{2.5}
\begin{array}{l}
\sigma({\rm Signal~1})\approx \sigma(pp\rightarrow \tilde{t}_{1}\tilde{t}_{1})\times {\rm BR}(\tilde{t}_{1}\rightarrow t\tilde{\chi}_{1}^{0})^{2} \\
 \sigma({\rm Signal~2})\approx \sigma(pp\rightarrow \tilde{t}_{1}\tilde{t}_{1})\times {\rm BR}(\tilde{t}_{1}\rightarrow b\tilde{\chi}_{1}^{\pm})^{2}\times{\rm BR}(\tilde{\chi}_{1}^{\pm}\rightarrow W^{\pm}\tilde{\chi}_{1}^{0})^{2} \\
 \sigma({\rm Signal~3})\approx \sigma(pp\rightarrow \tilde{t}_{1}\tilde{t}_{1})\times {\rm BR}(\tilde{t}_{1}\rightarrow b\tilde{\chi}_{1}^{\pm})^{2}\times{\rm BR}(\tilde{\chi}_{1}^{\pm}\rightarrow q\bar{q}^{\prime}\tilde{\chi}_{1}^{0})^{2}
\end{array}
\label{eq:approx}
\end{equation}
where the production cross-section of a pair of stop quarks are calculated by MadGraph, while the branching ratios are used from SPheno. The full matrix element calculation has been performed over a control group, and the approximation given in Eq.(\ref{eq:approx}) yields only about $0.7\%$ error in comparison to the full calculation \cite{Altin:2019veq}. The signal strength ($SS$) over the background processes is quantified as 

\begin{equation}
SS=\dfrac{S}{\sqrt{S+B}}~,
\end{equation}
where $S$ and $B$ refer to the event numbers (cross-section $\times$ Luminosity) of the signal and background respectively. We use the following correspondence to translate $SS$ to the confidence level (CL) \cite{Cranmer:2015nia}:

\begin{equation}
\setstretch{1.5}
\begin{array}{l}
0 \leq SS < 1 \rightarrow {\rm hardly~excluded}~, \\
1 \leq SS < 2 \rightarrow {\rm excluded~up~to~68\%}~, \\
2 \leq SS < 3 \rightarrow {\rm excluded~up~to~95\%}~.
\end{array}
\end{equation}

\section{Mass Spectrum and Signal Profile}
\label{sec:mass}

\begin{figure}[t!]
\centering
\subfigure{\includegraphics[scale=1.2]{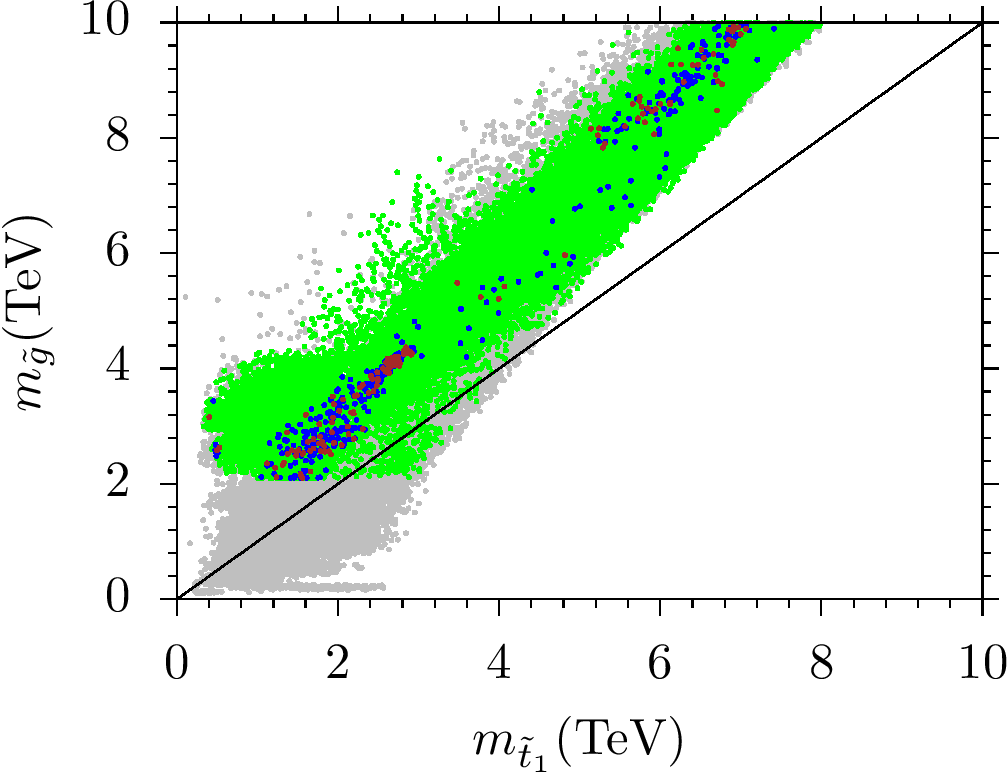}}%
\subfigure{\includegraphics[scale=1.2]{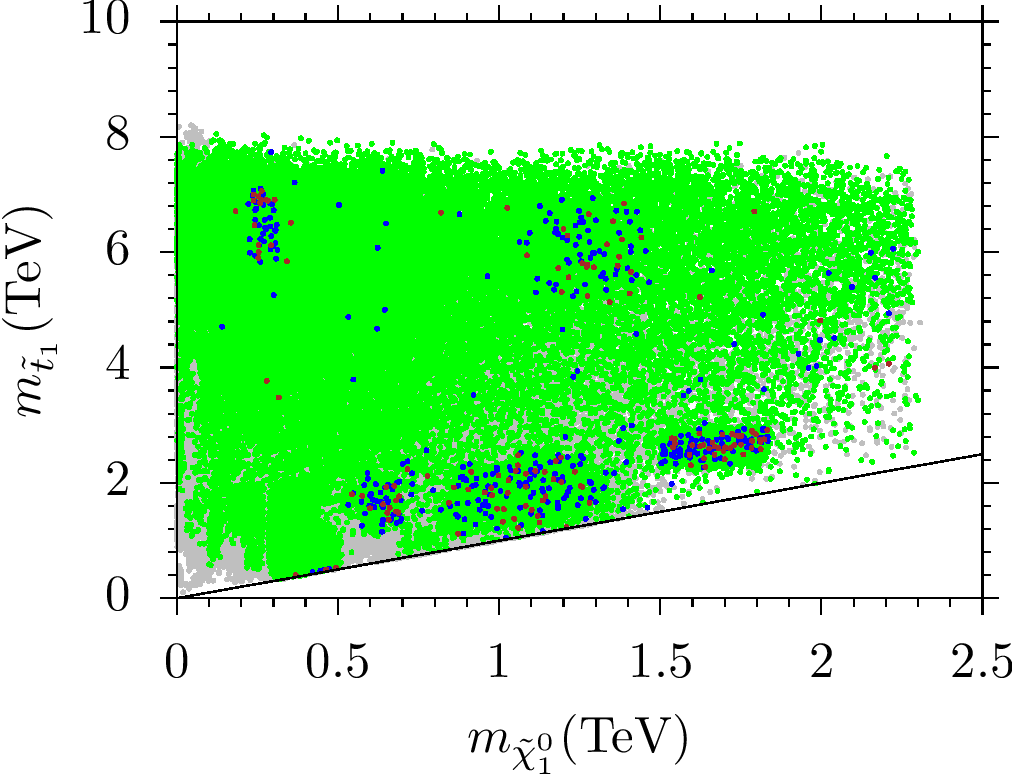}}
\caption{Plots in the $m_{\tilde{g}}-m_{\tilde{t}_{1}}$ and $m_{\tilde{t}_{1}}-m_{\tilde{\chi}_{1}^{0}}$ planes. All points are compatible with the REWSB and LSP neutralino conditions. Green points satisfy the mass bounds and the constraints from rare $B-$meson decays. Blue and red points form subsets of green, and they are allowed by the constraints on the relic abundance of LSP neutralino set by the WMAP and Planck satellites, respectively. The diagonal lines indicate the mass degeneracy between the displayed particles.}
\label{fig1}
\end{figure}

In this section we discuss the mass spectrum of SUSY particles with the emphasis on those, which participate in possible stop signals, and consider if the signal processes summarized in the previous section are available. Figure \ref{fig1} displays first the gluino and LSP neutralino masses in compared to the stop mass with plots in the $m_{\tilde{g}}-m_{\tilde{t}_{1}}$ and $m_{\tilde{t}_{1}}-m_{\tilde{\chi}_{1}^{0}}$ planes. All points are compatible with the REWSB and LSP neutralino conditions. Green points satisfy the mass bounds and the constraints from rare $B-$meson decays. Blue and red points form subsets of green, and they are allowed by the constraints on the relic abundance of LSP neutralino set by the WMAP and Planck satellites, respectively. The diagonal lines indicate the mass degeneracy between the particles displayed in the plots. The stop can be as heavy as about 8 TeV, while the gluino mass can go up to about 10 TeV and beyond, as seen in the $m_{\tilde{g}}-m_{\tilde{t}_{1}}$ plane. We did not show the region where $m_{\tilde{g}} > 10$ TeV, since it is beyond the reach of the current and future collider experiments, even after a high luminosity is achieved \cite{Altin:2019veq}. The results also reveal that the region with gluino lighter than stop (below the diagonal line in the $m_{\tilde{g}}-m_{\tilde{t}_{1}}$ plane) is mostly excluded by the current bound on the gluino mass, and only a small portion of the parameter space can kinematically  allow the $\tilde{t}\rightarrow \tilde{g}t$ process, which may not provide enough statistics to probe the stop through this channel. On the other hand, the $m_{\tilde{t}_{1}}-m_{\tilde{\chi}_{1}^{0}}$ plane shows that most of the solutions can yield enough mass difference between the stop and LSP neutralino that the $\tilde{t}_{1}\rightarrow t\tilde{\chi}_{1}^{0}$ process is allowed. The solutions around the diagonal line indicates the mass degeneracy between the stop and LSP neutralino. In this case, only the $\tilde{t}_{1}\rightarrow c \tilde{\chi}_{1}^{0}$ process is available. Even though the stop can be probe up to about 230 GeV through this decay channel, the DM constraints (blue and red points) already exclude the solutions with $m_{\tilde{t}_{1}}\lesssim 300$ GeV.

\begin{figure}[t!]
\centering
\subfigure{\includegraphics[scale=1.2]{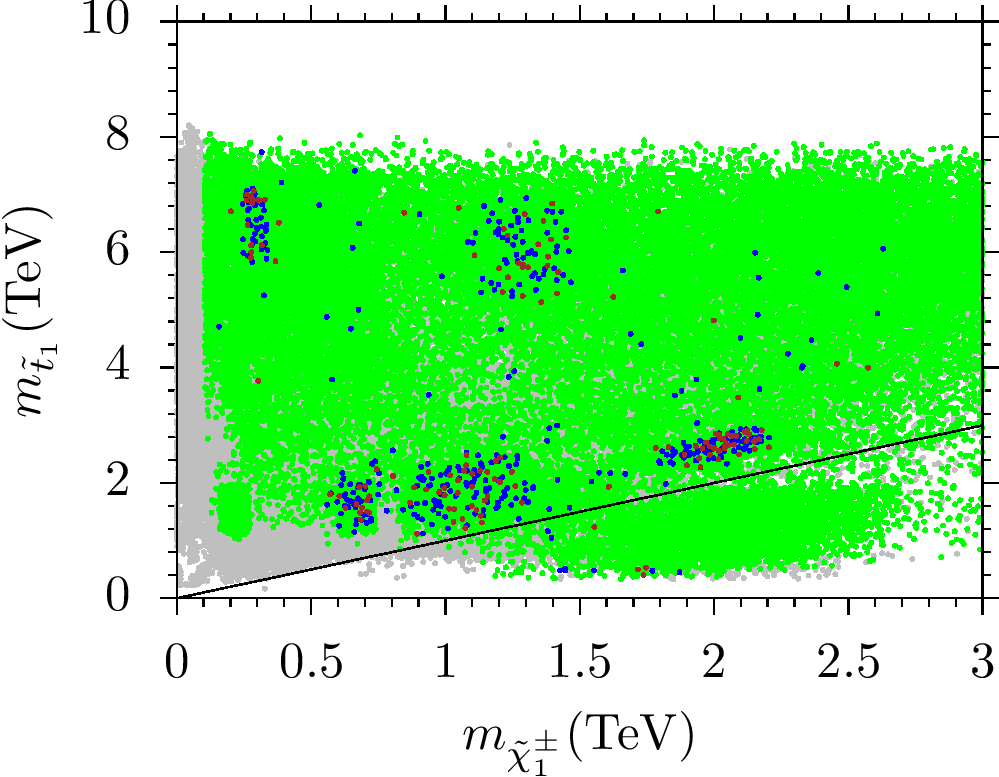}}%
\subfigure{\includegraphics[scale=1.2]{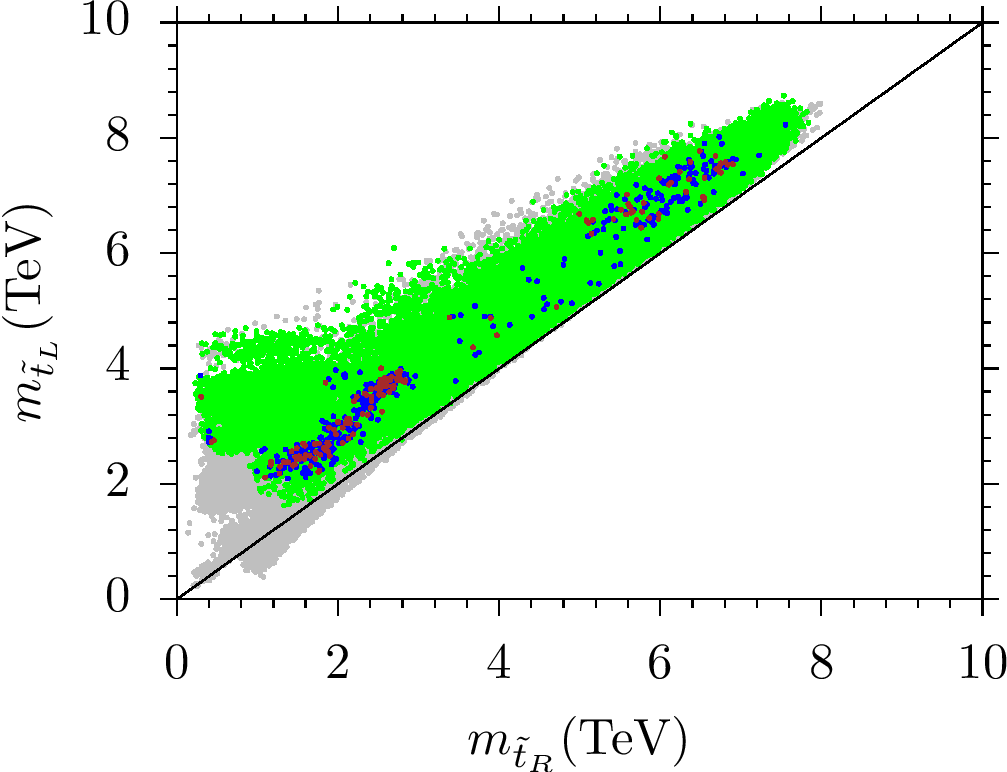}}
\caption{Plots in the $m_{\tilde{t}_{1}}-m_{\tilde{\chi}_{1}^{\pm}}$ and $m_{\tilde{t}_{L}}-m_{\tilde{t}_{R}}$ planes. The color coding is the same as in Figure \ref{fig1}.}
\label{fig2}
\end{figure}


We display our results in Figure \ref{fig2} with plots in the $m_{\tilde{t}_{1}}-m_{\tilde{\chi}_{1}^{\pm}}$ and $m_{\tilde{t}_{L}}-m_{\tilde{t}_{R}}$ planes. The color coding is the same as in Figure \ref{fig1}. The $m_{\tilde{t}_{1}}-m_{\tilde{\chi}_{1}^{\pm}}$ plane shows that the stop decay into a chargino is kinematically forbidden only in a small region (below the diagonal line), while their masses allow the process in most of the parameter space. However, the LHC allowed region yield the left-handed stop to be heavier than the right-handed state at all. The solutions around the diagonal line predict $m_{\tilde{t}_{L}}\gtrsim m_{\tilde{t}_{R}}$. In this case, the left-handed stop can provide a minor contribution to formation of the lightest chargino through its mixing with the right-handed stop. As is discussed in the previous section, when the lightest stop mass state is formed mostly by the right-handed stop, the processes involving $W^{\pm}-$boson, as in Signal 2, rather yield weak signal strengths, and one may not probe the stop through such events. Recall that Signal 3 can still be available, when the right-handed stop happens to be lighter than the left-handed stop.

\begin{figure}[t!]
\centering
\subfigure{\includegraphics[scale=1.2]{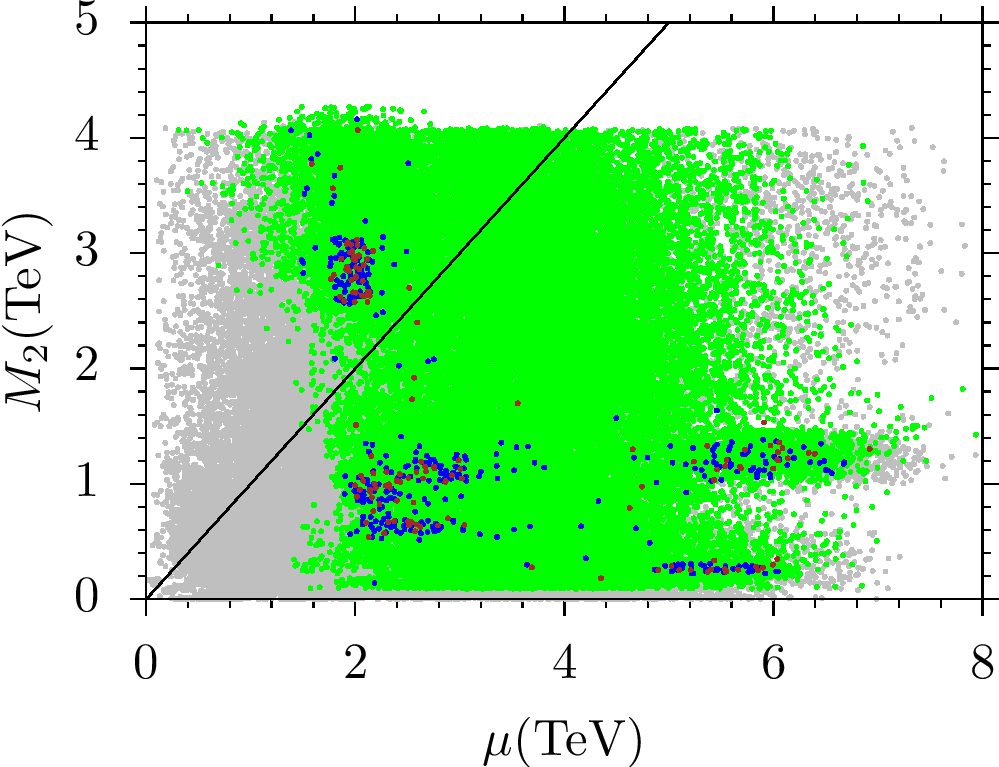}}%
\subfigure{\includegraphics[scale=1.2]{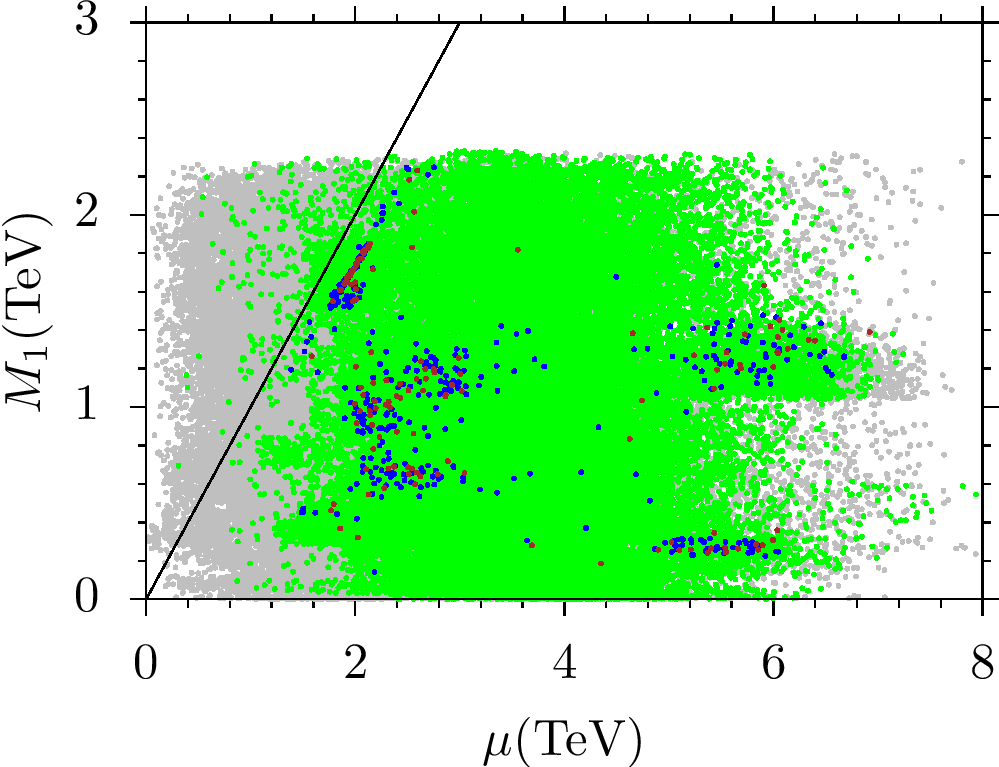}}
\caption{Plots in the $M_{2}-\mu$ and $M_{1}-\mu$ planes. The color coding is the same as in Figure \ref{fig1}}.
\label{fig3}
\end{figure}

In addition to the stop, Figure \ref{fig3} discusses formations of the chargino and LSP neutralino with plots in the $M_{2}-\mu$ and $M_{1}-\mu$ planes. The color coding is the same as in Figure \ref{fig1}. $M_{1}$ and $M_{2}$ represent the bino and wino masses at the SUSY scale, while $\mu$ stands for the Higgsino mass. As seen from the $M_{2}-\mu$ plane, Higgsino is realized heavier than wino over most of the parameter space. In this case, the lightest chargino is mostly formed by wino, which yields a weak signal strength for the $\tilde{t}_{1}\rightarrow b\tilde{\chi}_{1}^{\pm}$. However, there is a region above the diagonal line, in which the Higgsino happens to be lighter and forms the lightest chargino. In this case, even though ${\rm BR}(\tilde{t}_{1}\rightarrow b\tilde{\chi}_{1}^{\pm})$ can be large, Signal 2 is strongly suppressed since $m_{\tilde{\chi}_{1}^{\pm}}\sim m_{\tilde{\chi}_{1}^{0}}\sim \mu$ and ${\rm BR}(\tilde{\chi}_{1}^{\pm}\rightarrow W^{\pm}\tilde{\chi}_{1}^{0})\sim 0$. However, the processes of Signal 2 can be made available by replacing $\tilde{\chi}_{1}^{\pm} \rightarrow W^{\pm}\tilde{\chi}_{1}^{0}$ with $\tilde{\chi}_{1}^{\pm}\rightarrow q\bar{q}^{\prime}\tilde{\chi}_{1}^{0}$, as is classified in Signal 3. The $M_{1}-\mu$ plane shows that bino is mostly lighter than Higgsino, and comparing two panels of Figure \ref{fig3} reveals that the lightest neutralino is formed mostly by bino or bino-wino mixture, though bino-higgsino mixture is also available.

\section{Probing Stop in Collider Experiments}
\label{sec:stopprobe}

\begin{figure}[t!]
\centering
\subfigure{\includegraphics[scale=1.2]{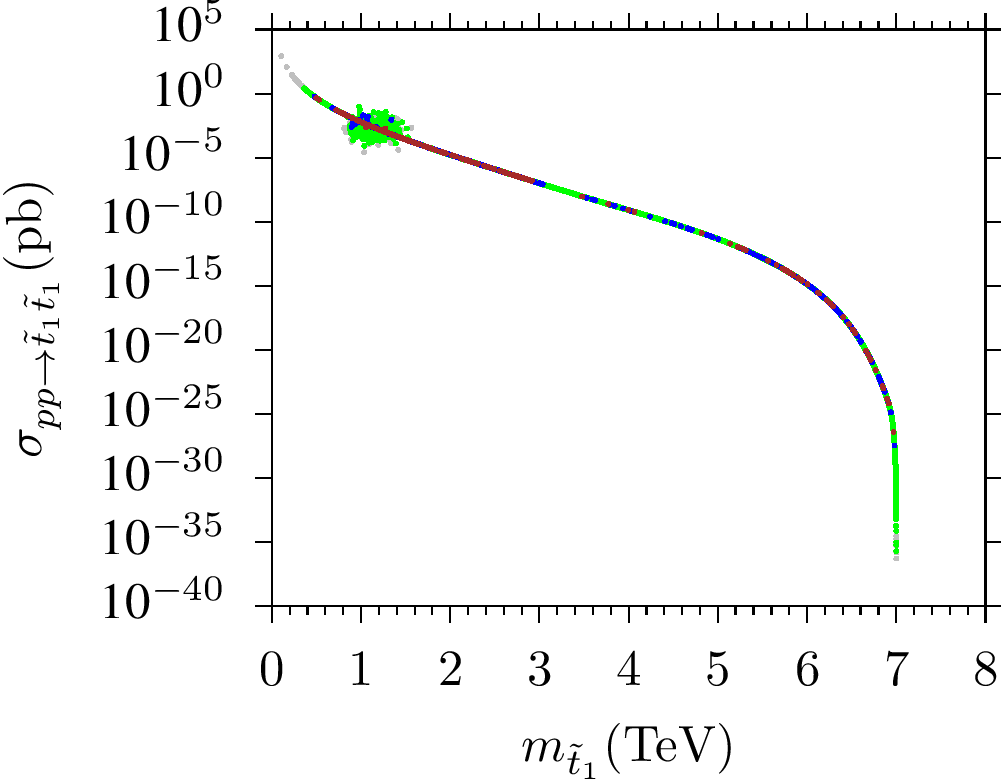}}%
\subfigure{\includegraphics[scale=1.2]{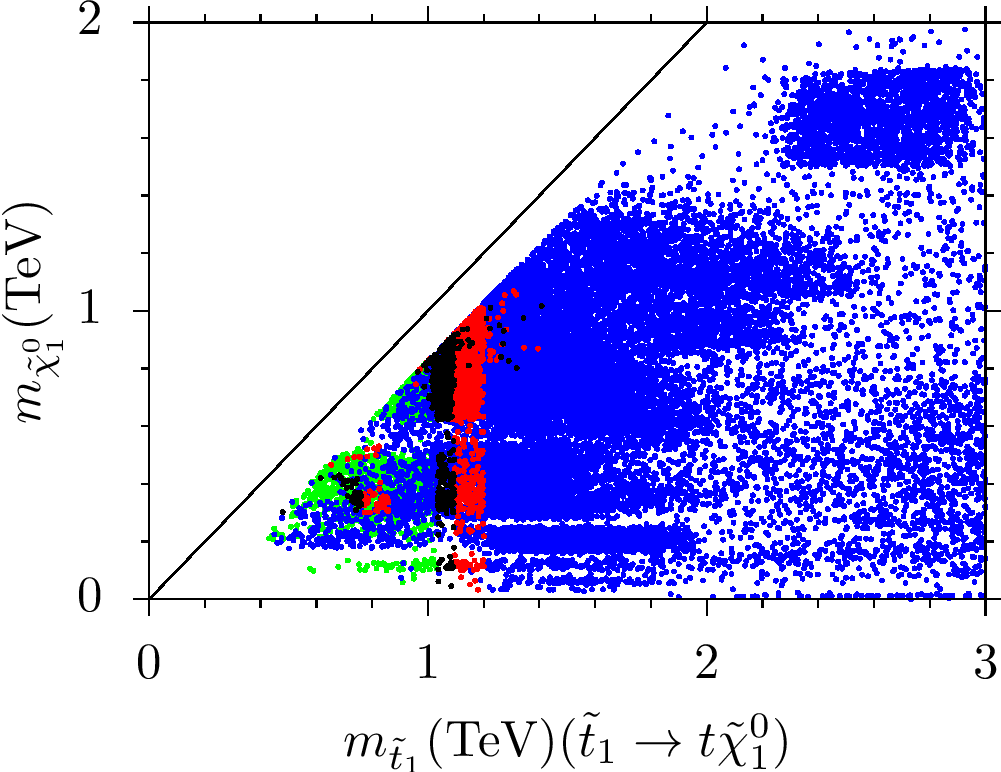}}
\caption{Plots for the stop pair production and LSP neutralino-stop mass correlation in the $\sigma(pp\rightarrow \tilde{t}_{1}\tilde{t}_{1})-m_{\tilde{t}_{1}}$ and $m_{\tilde{\chi}_{1}^{0}}-m_{\tilde{t}_{1}}$ planes. The color coding in the left panel is the same as in Figure \ref{fig1}. In the right panel, all solutions satisfy the mass bounds and the constraints from rare $B-$meson decays. Blue points represent the solutions with $SS({\rm Signal~1}) < 1$, while red, black and green correspond to $ 1 \leq SS({\rm Signal~1}) < 2$, $ 2 \leq SS({\rm Signal~1}) < 3$ and $SS({\rm Signal~1}) \geq 3$, respectively for 14 TeV with 36.1 fb$^{-1}$ luminosity.}
\label{fig4}
\end{figure}

\begin{figure}[t!]
\subfigure[27 TeV, 36.1 fb$^{-1}$]{\includegraphics[scale=1.2]{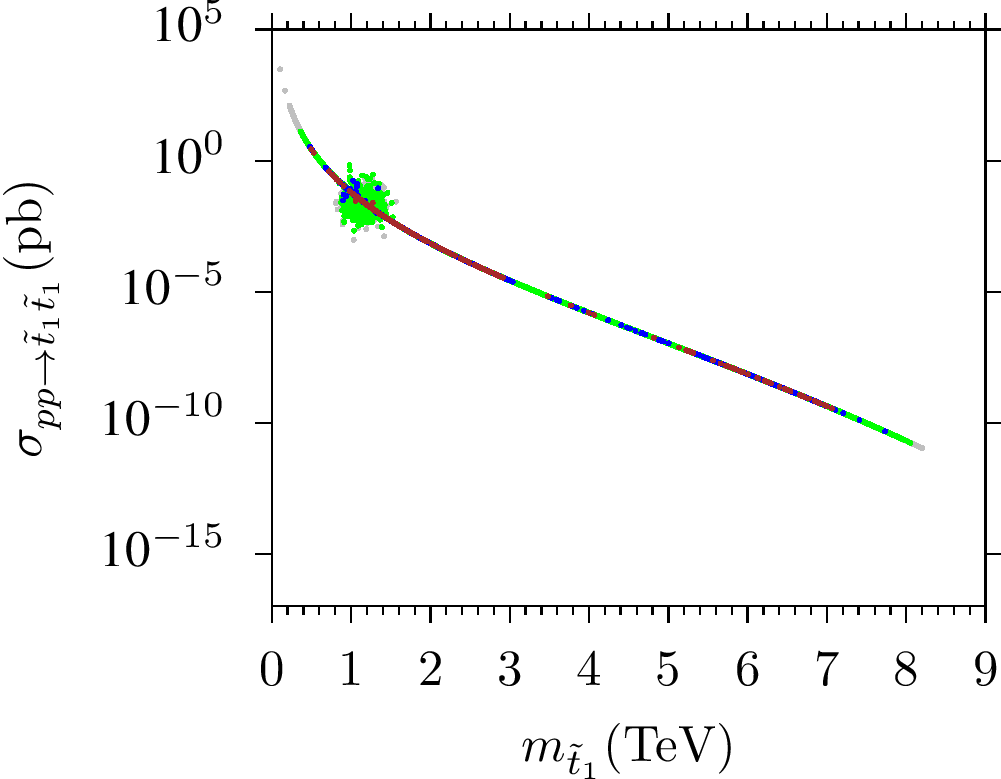}}%
\subfigure[100 TeV, 36.1 fb$^{-1}$]{\includegraphics[scale=1.2]{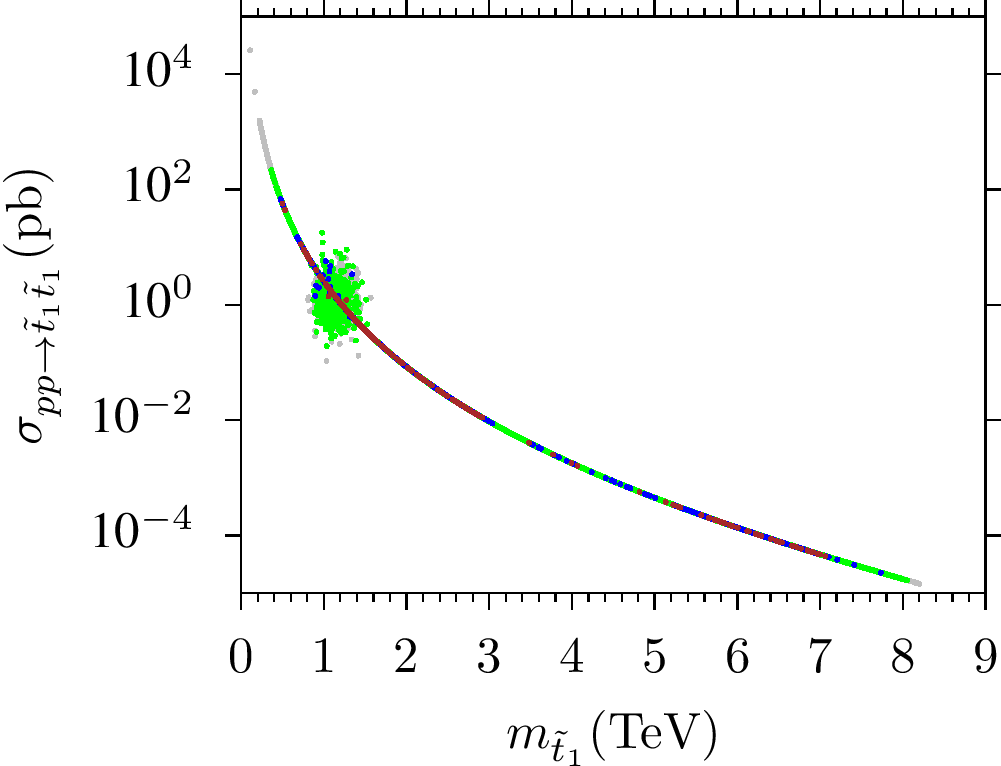}}
\subfigure{\hspace{1.0cm} \includegraphics[scale=1]{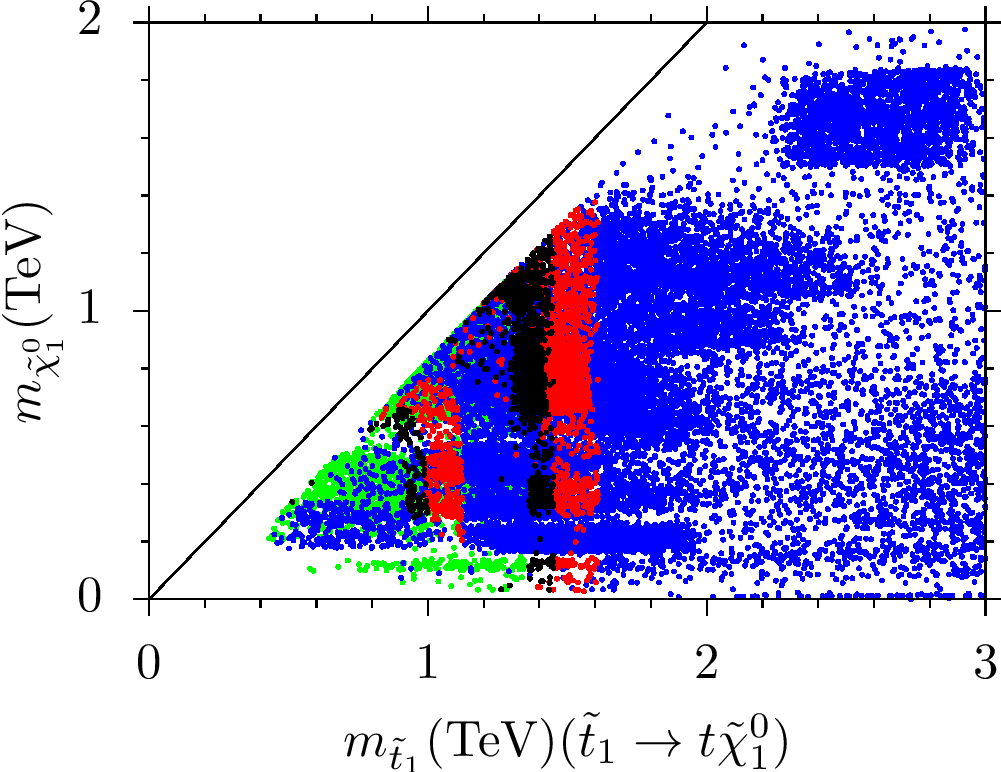}}%
\subfigure{\hspace{1.0cm}\includegraphics[scale=1.05]{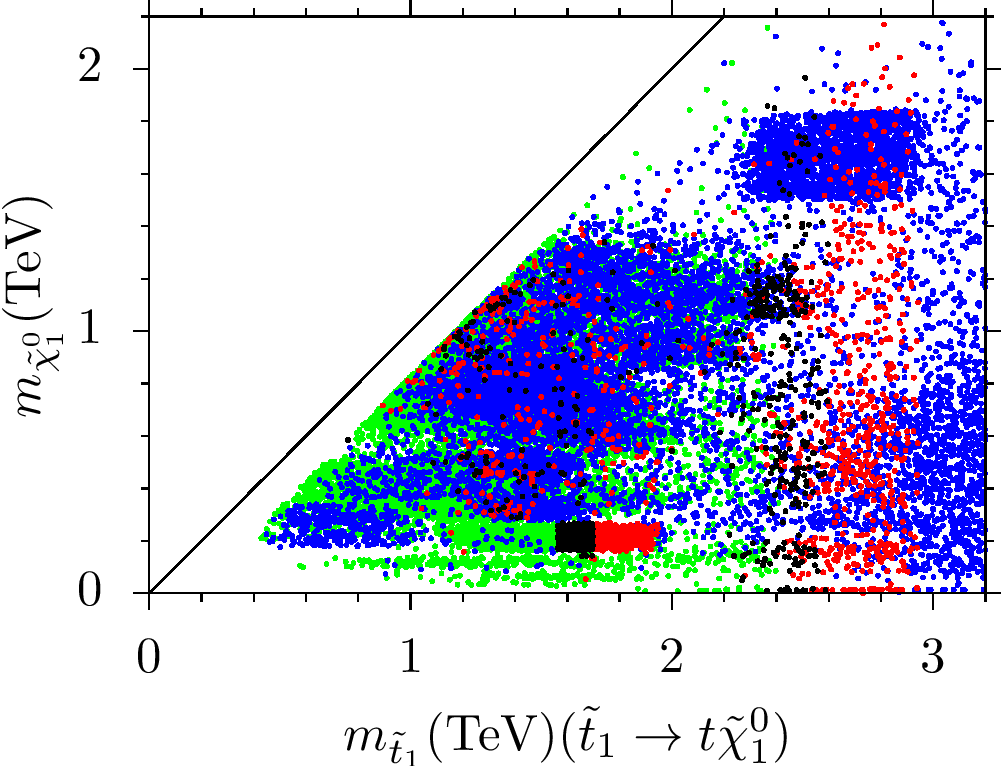}}
\caption{Plots for the stop pair production and LSP neutralino-stop mass correlation in the $\sigma(pp\rightarrow \tilde{t}_{1}\tilde{t}_{1})-m_{\tilde{t}_{1}}$ and $m_{\tilde{\chi}_{1}^{0}}-m_{\tilde{t}_{1}}$ planes at 27 TeV (left) and 100 TeV (right) center of mass energies. The color coding is the same as in Figure \ref{fig4}.}
\label{fig5}
\end{figure}

In the previous section, we discussed the available signal processes in terms of masses and flavors of the relevant particles, and concluded that $\tilde{t}_{1}\rightarrow t\tilde{\chi}_{1}^{0}$ and the $\tilde{t}_{1}\rightarrow b\tilde{\chi}_{1}^{\pm}\rightarrow bq\bar{q}^{\prime}\tilde{\chi}_{1}^{0}$ decay modes are likely to provide most promising signals. However, even though we select the most optimized benchmark points which yield the largest branching ratios in the processes listed in Eq.(\ref{eq:sigs}), the main suppression comes from the smallness of the stop pair production cross-section. Figure \ref{fig4} represents the cross-section for the stop pair production and the strength of the processes classified in Signal 1 with plots in the $\sigma(pp\rightarrow \tilde{t}_{1}\tilde{t}_{1})-m_{\tilde{t}_{1}}$ and $m_{\tilde{\chi}_{1}^{0}}-m_{\tilde{t}_{1}}$ planes. The color coding in the left panel is the same as in Figure \ref{fig1}. In the right panel, all solutions satisfy the mass bounds and the constraints from rare $B-$meson decays. Blue points represent the solutions with $SS({\rm Signal~1}) < 1$, while red, black and green correspond to $ 1 \leq SS({\rm Signal~1}) < 2$, $ 2 \leq SS({\rm Signal~1}) < 3$ and $SS({\rm Signal~1}) \geq 3$, respectively for 14 TeV with 36.1 fb$^{-1}$ luminosity. The $\sigma(pp\rightarrow \tilde{t}_{1}\tilde{t}_{1})-m_{\tilde{t}_{1}}$ plane shows that the stop pair in the LHC allowed region (green) can be produced at $\sigma(pp\rightarrow \tilde{t}_{1}\tilde{t}_{1}) \sim 1$ pb when the stop weighs about 500 GeV, and it drops below $10^{-5}$ pb for $m_{\tilde{t}_{1}} \gtrsim 2$ TeV. Even though the correlation between the cross-section of the stop-pair production and the stop mass is mostly a smooth curve, there is a region of scattered points. This region happens as a result of an interplay among the different processes of stop pair-production such as $q\bar{q}$ interactions and hard gluon corrections, as mentioned before. Following the approximation given in Eq.(\ref{eq:approx}) and assuming absence of a direct signal, the $m_{\tilde{\chi}_{1}^{0}}-m_{\tilde{t}_{1}}$ plane shows that the results can exclude the stop masses up to about 1.2 TeV at $68\%$ CL (red), while $m_{\tilde{t}_{1}}\lesssim 1.1$ TeV is excluded at $95\%$ CL (black). Note that the blue points represent the solutions with $SS < 1$, and the analyses do not apply to these solutions.

We continue to discuss the stop pair production and the strength of Signal 1 in Figure \ref{fig5} with plots in the $\sigma(pp\rightarrow \tilde{t}_{1}\tilde{t}_{1})-m_{\tilde{t}_{1}}$ and $m_{\tilde{\chi}_{1}^{0}}-m_{\tilde{t}_{1}}$ planes at 27 TeV (left) and 100 TeV (right) center of mass energies. The color coding is the same as in Figure \ref{fig4}. The luminosity is set to 36.1 fb$^{-1}$ in both cases. When the center of mass energy is raised to 27 TeV, the stop pair production cross-section is realized of the order $\mathcal{O}(10)$ pb, while it can be expected as high as $ \mathcal{O}(100)$ pb in the collider experiments with 100 TeV center of mass energy, as seen from the top panels. The strength of Signal 1 should be expected to raise accordingly. The bottom panel shows that the stop can be probed to about 1.6 TeV at 27 TeV, and about 3 TeV at 100 TeV with $68\%$ CL (red). Requiring $95\%$ CL yield the mass scales of about 1.4 TeV at 27 TeV, and 2.6 TeV at 100 TeV (black) to probe the stop.

\begin{figure}[t!]
\centering
\subfigure{\includegraphics[scale=1.2]{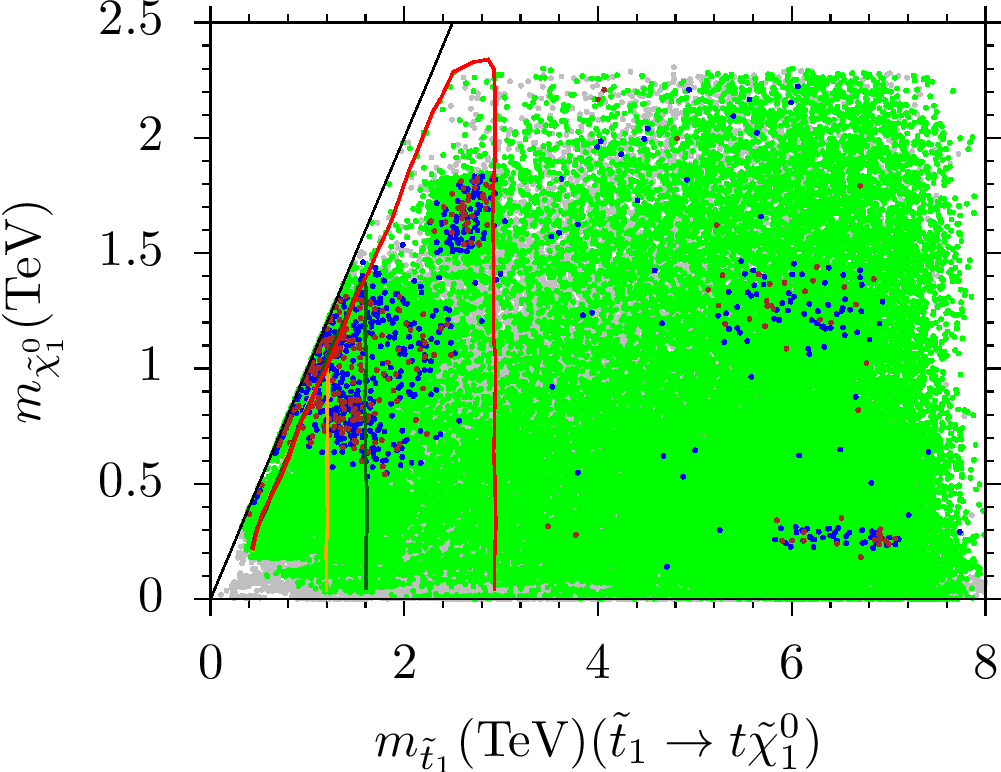}}%
\subfigure{\includegraphics[scale=1.2]{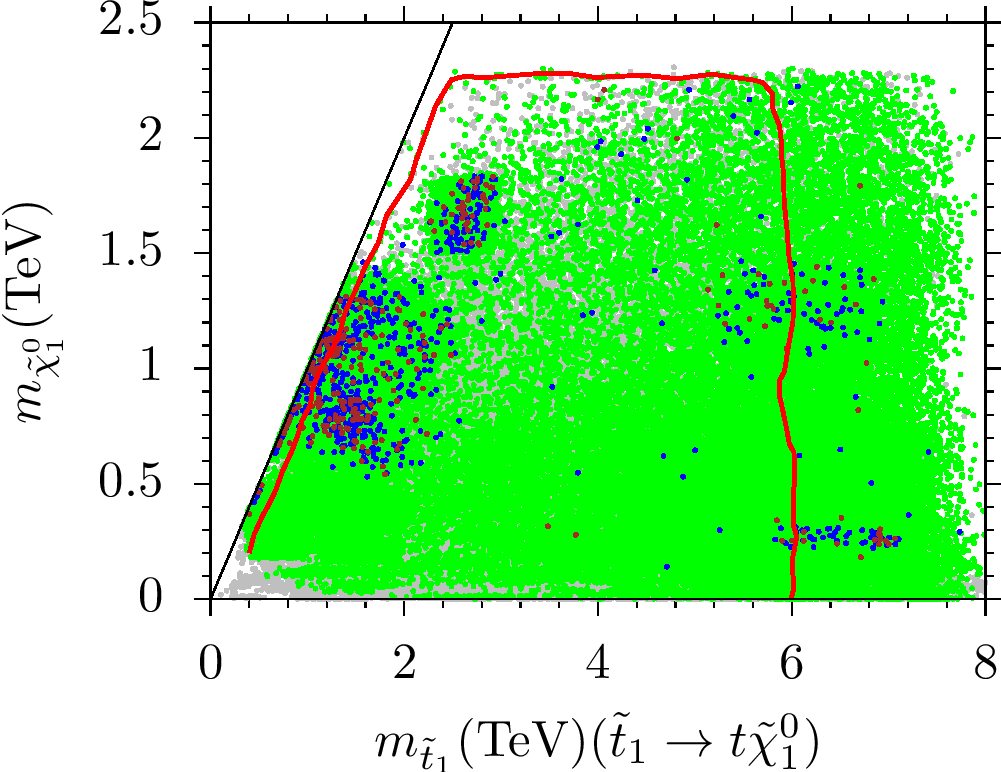}}
\caption{Exclusion curves for the stop mass through $\tilde{t}_{1}\rightarrow t\tilde{\chi}_{1}^{0}$. The color coding is the same as in Figure \ref{fig1}. In the left panel, orange curve represents the exclusion at the current energy, while the dark green and red are obtained at 27 TeV and 100 TeV, repsectively. Luminosity is set to 36.1 fb$^{-1}$ at all energies. The right panel is the same exclusion curve for the stop mass at 100 TeV center of mass energy and 3000 fb$^{-1}$ luminosity.}
\label{fig6}
\end{figure}

Figure \ref{fig6} summarizes our finding in the analyses over Signal 1 in the $m_{\tilde{\chi}_{1}^{0}}-m_{\tilde{t}_{1}}$ plane. The color coding is the same as in Figure \ref{fig1}. In the left panel, orange curve represents the exclusion at the current energy, while the dark green and red are obtained at 27 TeV and 100 TeV, repsectively. Luminosity is set to 36.1 fb$^{-1}$ at all energies. The right panel is the same exclusion curve for the stop mass at 100 TeV center of mass energy and 3000 fb$^{-1}$ luminosity. As seen from the left panel, the current collider experiments can exclude the stop within the mass scales up to about 1.2 TeV, while the future experiments with 27 TeV and 100 TeV center of mass energy are promising to probe the stop up to about 1.6 TeV and 3 TeV, respectively. The right panel displays the reachable mass scales when 3000 fb$^{-1}$ luminosity is achieved at 100 TeV. As seen from the results, the stop will be able to be probed up to about 6 TeV, if it decays into a LSP neutralino along with a top quark.

\begin{figure}[t!]
\centering
\subfigure[14 TeV, 36.1 fb$^{-1}$]{\includegraphics[scale=1.2]{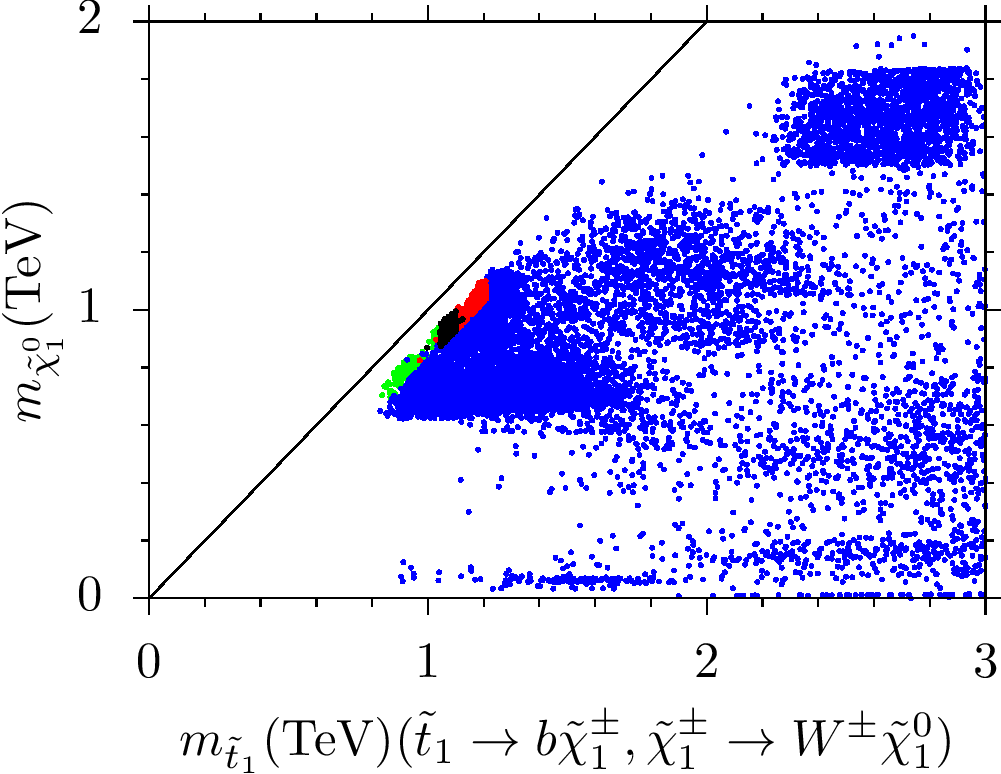}}%
\subfigure[100 TeV, 3000 fb$^{-1}$]{\includegraphics[scale=1.2]{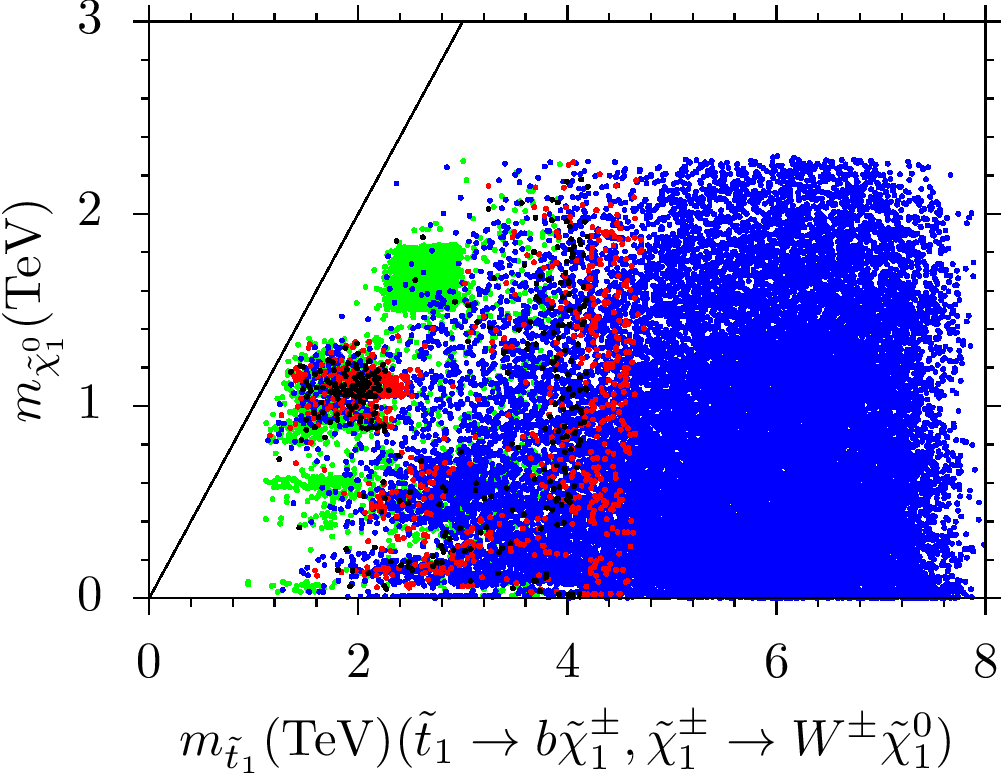}}
\caption{Plots in the $m_{\tilde{\chi}_{1}^{0}}-m_{\tilde{t}_{1}}$ plane at 14 TeV with 36.1 fb$^{-1}$ luminosity (left), and 100 TeV with 3000 fb$^{-1}$ luminosity (right). All solutions satisfy the mass bounds and the constraints from rare $B-$meson decays. Blue points represent the solutions with $SS({\rm Signal~2}) < 1$, while red and black correspond to $ 1 \leq SS({\rm Signal~2})\leq 2$, and $ 2 \leq SS({\rm Signal~2})\leq 3$, respectively.}
\label{fig7}
\end{figure}

As we discussed in the previous section, Signal 2 is not expected to be strong enough due to the right-handed stop forming the lightest stop mass eigenstate, unless the left-handed stop can take part considerably in formation of $\tilde{t}_{1}$. Such solutions can be realized when $m_{\tilde{t}_{L}}\gtrsim m_{\tilde{t}_{R}}$. However, such solutions were realized in a small region, and their impact is shown in Figure \ref{fig7} with plots in the $m_{\tilde{\chi}_{1}^{0}}-m_{\tilde{t}_{1}}$ plane at 14 TeV with 36.1 fb$^{-1}$ luminosity (left), and 100 TeV with 3000 fb$^{-1}$ luminosity (right). All solutions satisfy the mass bounds and the constraints from rare $B-$meson decays. Blue points represent the solutions with $SS({\rm Signal~2}) < 1$, while red and black correspond to $ 1 \leq SS({\rm Signal~2})\leq 2$, and $ 2 \leq SS({\rm Signal~2})\leq 3$, respectively. The green, black and red points indicate that Signal 2 can be available only if the solutions lead to $0.8 \lesssim m_{\tilde{t}_{1}}\lesssim 1.2$ TeV and $0.6 \lesssim m_{\tilde{\chi}_{1}^{0}}\lesssim 1$ TeV. On the other hand, the right panel displays interesting results that Signal 2 becomes available for the analyses when 3000 fb$^{{-1}}$ luminosity is reached at 100 TeV collider experiments. Indeed, one can probe the stop mass up to about 4.8 TeV through Signal 2.

\begin{figure}[t!]
\centering
\subfigure{\includegraphics[scale=1.2]{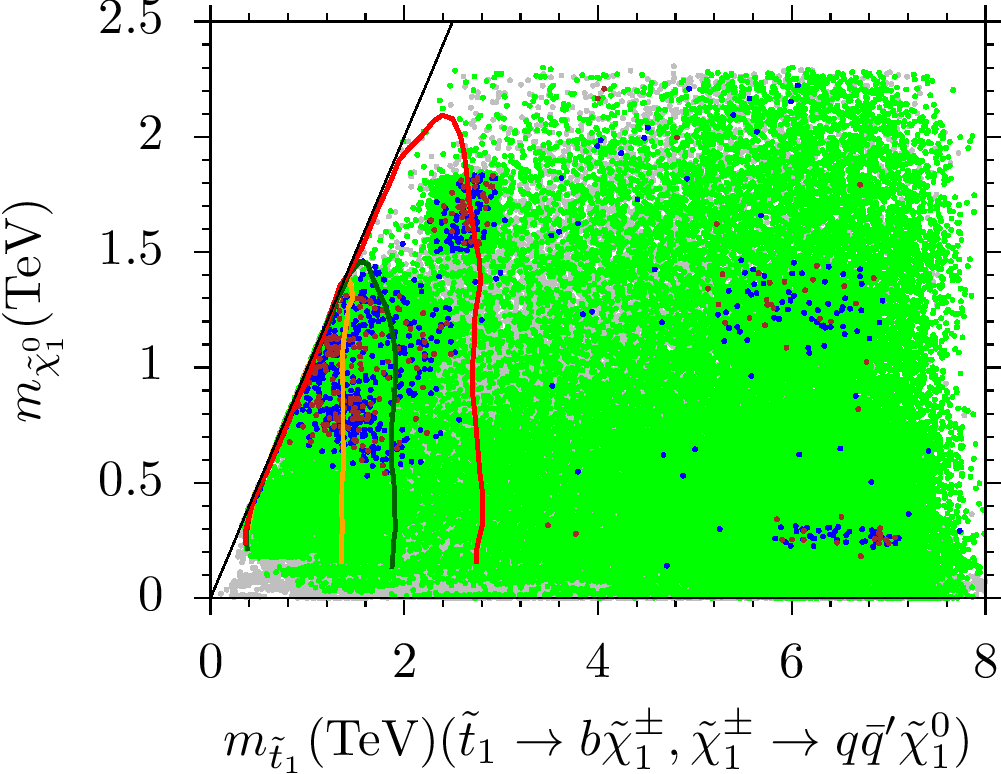}}%
\subfigure{\includegraphics[scale=1.2]{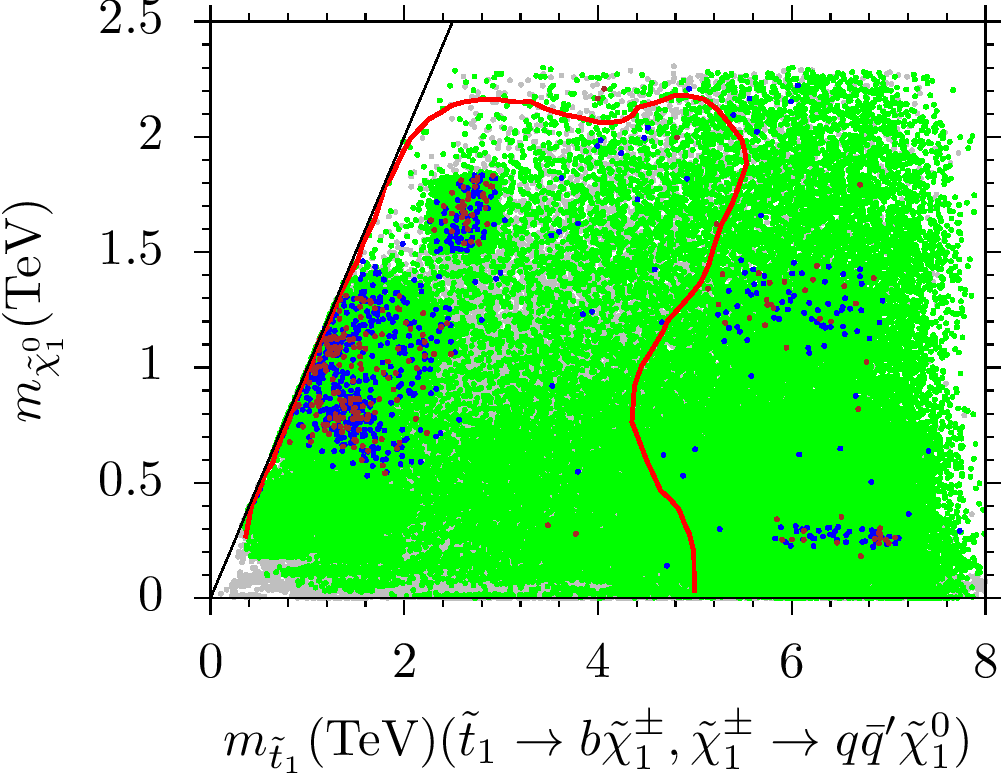}}
\caption{Exclusion curves for the stop mass through $\tilde{t}_{1}\rightarrow b\tilde{\chi}_{1}^{\pm}$ with $\tilde{\chi}_{1}^{\pm}\rightarrow q\bar{q}^{\prime}\tilde{\chi}_{1}^{0}$. The color coding is the same as in Figure \ref{fig1}. In the left panel, orange represents the exclusion at the current energy, while the dark green and red are obtained at 27 TeV and 100 TeV, repsectively. Luminosity is set to 36.1 fb$^{-1}$ at all energies. The right panel is the same exclusion curve for the stop mass at 100 TeV center of mass energy and 3000 fb$^{-1}$ luminosity. }
\label{fig8}
\end{figure}

Finally, we discuss the results for Signal 3 from similar analyses performed for Signal 1 and Signal 2 in Figure \ref{fig8} by displaying the exclusion curves at various center of mass energies. The color coding is the same as in Figure \ref{fig1}. In the left panel, orange represents the exclusion at the current energy, while the dark green and red are obtained at 27 TeV and 100 TeV, repsectively. Luminosity is set to 36.1 fb$^{-1}$ at all energies. The right panel is the same exclusion curve for the stop mass at 100 TeV center of mass energy and 3000 fb$^{-1}$ luminosity. The left panel shows that the current experiments can exclude the stop mass below about 1.2 TeV, while it can be probed up to about 2 TeV at 27 TeV and 2.8 TeV at 100 TeV. The right panel reveals that the future experiments can probe the stop masses further up to about 5 TeV through Signal 3.

\section{Conclusion}
\label{sec:conc}
We have discussed the stop masses and possible signal process within a class of SUSY GUTs with non-universal gaugino masses. This class of models predicts the stop mass in a wide range  from about 400 GeV to 8 TeV, and the DM constraints yield a further lower bound on the stop mass as $m_{\tilde{t}_{1}}\gtrsim 500$ GeV. The mass spectrum also includes the gluino mass from 2.1 TeV (as stated by the gluino mass bound) to beyond 10 TeV. It usually happens to be heavier the stop, despite the presence of a very narrow region in which the stop can slightly be heavier than gluino. However, this region is excluded by the DM constraints. Being the LSP neutralino always takes part in possible signal processes, and its mass is realized as heavy as about 2.3 TeV in the fundamental parameter space. Even though LHC allowed region allows almost massless neutralino, the DM constraints bounds its mass at about 200 GeV from below. Similarly the lightest chargino mass can be realized beyond 3 TeV, while the DM constraints bound its mass at about 2.7 TeV from above.

We consider three possible signal processes to probe the stop in this class of SUSY GUTs. The strongest impact can be obtained when a stop significantly decays into a LSP neutralino along with a top quark, which refers to Signal 1. The current experiments at 14 TeV center of mass energy with 36.1 fb$^{-1}$ luminosity can exclude the stop mass up to about 1.2 TeV through this decay channel, while the stop mass can be expected to be probed up to about 1.6 TeV in the experiments with 27 TeV center of mass energy, while the probe scale will raise to about 3 TeV when the 100 TeV center of mass energy is set in the collider experiments. If 3000 fb$^{-1}$ luminosity is reached in 100 TeV collider experiments, this channel can probe the stop mass up to about 6 TeV. Another decay mode of the stop can be listed as Signal 2, in which it decays into a bottom quark along with a chargino, and the chargino successively decays into a $W-$boson and LSP neutralino. We found that this channel is not available for the current collider experiments, since the lightest stop mass eigenstate is formed mostly by the right-handed stop. Only a small region with $0.8 \lesssim m_{\tilde{t}_{1}} \lesssim 1.2$ TeV can yield a considerable left-handed stop mixing, and it leads to visible signal. Despite the lack of good statistics for the current experiments, this channel can be expected to be available and probe the stop mass to about 4.8 TeV in the experiments with 100 TeV center of mass energy and 3000 fb$^{-1}$ luminosity. However, a similar signal can be considered in which the chargino's successive decay is into two quarks and a LSP neutralino, which is stated as Signal 3 in our discussion. Similar analyses have revealed that this channel is available for the current experiments, and it excludes the solutions with $m_{\tilde{t}_{1}}\lesssim 1.2$. It provides also a promising signal through which the stop mass can be probed to about 2 TeV at 27 TeV, and 2.8 TeV at 100 TeV center of mass energies. Besides, high luminosity raises the probing scale for the stop mass up to about 5 TeV.

\section{Acknowledgments}

The work of Z. A, Z. K. and C. S. U is supported by the Scientific and Technological Research Council of Turkey (TUBITAK) Grant no. MFAG-118F090. Part of the calculations reported in this paper were performed at the National Academic Network and Information Center (ULAKBIM) of TUBITAK, High Performance and Grid Computing Center (TRUBA Resources)

\end{document}